\documentclass[12pt,english,preprint,superscriptaddress]{revtex4}
\usepackage[latin1]{inputenc}
\usepackage[english]{babel}
\usepackage[T1]{fontenc}
\usepackage[dvips]{graphicx}
\usepackage{color}
\usepackage{subfigure}
\date{\today}
\begin{document}
 
 \title{Ultracold dipolar few-boson ensembles in a triple well trap}

\author{Budhaditya Chatterjee}
\email{bchatter@physnet.uni-hamburg.de}
\affiliation{Zentrum f\"ur Optische Quantentechnologien,  Universit\"at Hamburg, Luruper Chaussee 149, 22761 Hamburg, Germany}

\author{Ioannis Brouzos}
\email{ibrouzos@physnet.uni-hamburg.de}
\affiliation{Zentrum f\"ur Optische Quantentechnologien,  Universit\"at Hamburg, Luruper Chaussee 149, 22761 Hamburg, Germany}

\author{Lushuai Cao}
\email{lcao@physnet.uni-hamburg.de}
\affiliation{Zentrum f\"ur Optische Quantentechnologien,  Universit\"at Hamburg,  Luruper Chaussee 149, 22761 Hamburg, Germany}
\affiliation{The Hamburg Centre for Ultrafast Imaging, Luruper Chaussee 149, 22761 Hamburg, Germany}

\author{Peter Schmelcher}
\email{pschmelc@physnet.uni-hamburg.de}
\affiliation{Zentrum f\"ur Optische Quantentechnologien,  Universit\"at Hamburg, Luruper Chaussee 149, 22761 Hamburg, Germany}
\affiliation{The Hamburg Centre for Ultrafast Imaging, Luruper Chaussee 149, 22761 Hamburg, Germany}
 
\begin{abstract}

We investigate the ground state properties and tunneling dynamics of ultracold dipolar bosons in a one dimensional triple well trap from a few-body ab-initio perspective. Our focus is primarily on  the distinctive features of dipolar bosons compared to the contact interacting bosons. Formation of intra-well localization is observed for very strong dipolar interaction. General population rearangement as well as fragmentation and localization effects have been found, depending strongly on the particle number. The energy spectrum for two particles exhibits avoided crossings that lead to several distinct resonances involving different bands, i.e. to an inter-band resonant tunneling dynamics. The corresponding mechanisms are investigated by studying among others the pair-probability and performing an eigenstate analysis.

 \end{abstract}

\maketitle

\section{Introduction}

Recently dipolar quantum gases\cite{baranov08,lahaye09}  have become a major focus in the field of ultracold atoms. The important experimental progress of achieving Bose-Einstein condensates of chromium atoms \cite{griesmaier05, beaufils08} and more recently of dysprosium atoms \cite{lu11}, as well as a high phase-space density gas of  polar KRb molecule \cite{ni08}  has strongly motivated these studies. The important feature of dipolar gases is their long range and anisotropic interaction which results in interesting effects such as an elongation of the condensate along the direction of the oriented dipoles \cite{yi01, santos00, goral00} or the stability of pure dipolar condensates in special geometries such as a very oblate trap \cite{yi01, santos00, eberlein05, goral02a,koch08}. Dipolar gases thus represent an important tool to study and simulate a variety of novel quantum effects and exotic quantum phases \cite{lahaye09, goral02, menotti07, zoller10}.Of special importance is the study of dipolar gases in 
lower dimensions which often results in new phases and effects not observed in three dimensions such as  p-wave Fermi superfluids in two dimensions \cite{bruun08, cooper09} and Luttinger-liquid-like behavior in one dimension  \cite{arkhipov05,citro07,depalo08,pedri08}. Moreover dipoles, polarized perpendicularly in one-dimension, interact with repulsive interaction thus alleviating the collisional instability arising from the head-to-tail alignment of the dipoles. 

Investigations of dipolar gases in one dimensional lattices  demonstrate interesting features not observed for bosons with  on-site interaction, such as density waves \cite{goral02, dalla06} and Haldane insulating phases \cite{dalla06, deng11}. Moreover, dipolar gases in a  ring geometry have been shown to  exhibit anisotropic interactions leading to the formation of a variety of new phases \cite{zollner11,maik11}. Dipolar bosons in a triple well potential have been studied in ref \cite{lahaye10,peter12} showing the non-local character of the interaction  as well as dynamical superpositions \cite{lahaye10}. Also in the case of the corresponding quantum dynamics, dipolar gases show a rich behavior as a consequence of their long-range nature. For instance, the tunneling dynamics of dipolar degenerate gases in double-well trap exhibits symmetry breaking and self-trapping \cite{xiong09} while interesting tunneling mechanisms have been reported for toroidal traps \cite{abad10}.  


In this work, we investigate the ground state properties and tunneling dynamics of ultracold few dipolar bosons in a one dimensional triple well trap from a bottom  up ab-initio perspective using the numerically exact Multi-Configuration Time Dependent Hartree (MCTDH) method \cite{meyer90,beck00}. The ground state properties along the crossover from weak to strong interactions are studied with a focus on microscopic quantum effects and mechanisms occurring in the system. We elaborate the differences of dipolarly versus contact interacting bosonic systems \cite{brouzos10}. Fragmentation and localization scenarios for very strong interactions as well as inter-well population realignment depending strongly on the filling fractions are derived. The energy spectrum for two particles show  avoided crossings which allow for interesting inter-band tunneling processes reminiscent of those for contact bosons in a triple well \cite{cao11}. 

The paper is organized as follows. In Section \ref{sec:setup} we introduce our model and the system set up. In Sec. \ref{sec:gs} we investigate the ground state for different particle numbers. The energy spectrum for two bosons is investigated in Sec. \ref{sec:spec} and subsequently in Sec. \ref{sec:dyn} we explore the inter-band tunneling dynamics for this case. Sec. \ref{sec:conclusion} contains the conclusions.

\section{Model and setup}
\label{sec:setup}

Our system consists of $N$ interacting dipolar bosons in a  $1D$ triple well trap. The general Hamiltonian can be written as 

\begin{equation}
{
H=\sum_{i=1}^{N}-\frac{\hbar^{2}}{2M}\partial_{x_{i}}^{2}+\sum_{i=1}^{N}V_{tw}(x_{i})+\frac{1}{2}\sum_{i\neq j}V_{int}(x_{i}-x_{j})
}.
\label{Eq.Ham}
\end{equation}

Here, the $V_{tw}$ represents the triple well potential which we model as $V_{tw}= V_{0}sin^{2}(\kappa x)$ (where $V_0$ gives the lattice depth and $\kappa$ is the wave vector). In order to confine the bosons in the triple well  we impose a hard wall boundary condition at $x=\pm3\pi/2\kappa$. 

The interaction potential $V_{int}$ consists in general of a combination of short ranged contact interaction  and the dipolar interaction. In order to realize pure dipolar effects, we omit the short-ranged contact interactions (they can be  experimentally tuned to zero using magnetic or optical Feshbach resonances \cite{chin}). The repulsive dipolar interaction between two bosons whose dipole moments are aligned by an electric field is given by $V_d(x_i -x_j) = D^2/|x_i - x_j|^3$ where $D^2 = d^2_m /4\pi\epsilon_0$ for electric dipoles  and $D^2 = d^2_m \mu_0 /4\pi$ for magnetic dipoles  ($d_m$ being the dipole moment). 

For reasons of universality as well as computational convenience, we will rescale Eq. \ref{Eq.Ham} in units of the recoil energy $E_{R}=\hbar^{2}\kappa^{2}/2M$ thereby setting $\hbar=M=\kappa=1$. The scaled triple well potential is given as $V'_{tw}=V_{0}sin^{2}x$  with hard wall boundary conditions at $x=\pm3\pi/2$. The scaled dipolar interaction strength becomes $D^2/\hbar^{2}\kappa = d$. The length and time units are given in terms of $\kappa^{-1}$ and $\hbar/E_{R}$, respectively. $d$ thus represents a dimensionless dipolar interaction strength. All quantities are in dimensionless units throughout.


\section{Ground State Properties}
\label{sec:gs}

\subsection{Commensurate Filling}
\label{sec:com_fill}

The ground state configuration is determined by the competition between the external trapping confinement and the interaction between the particles. In the case of  commensurate filling, the most important effect is the superfluid to Mott-insulator transition. For small interactions, the effects of dipolar interactions are almost identical to that of contact interactions and it is only for significantly stronger interaction strengths that the two  could be distinguished, in particular concerning on-site effects. 

\begin{figure}[htb]
\begin{center}
\includegraphics[width=0.65\columnwidth,keepaspectratio]{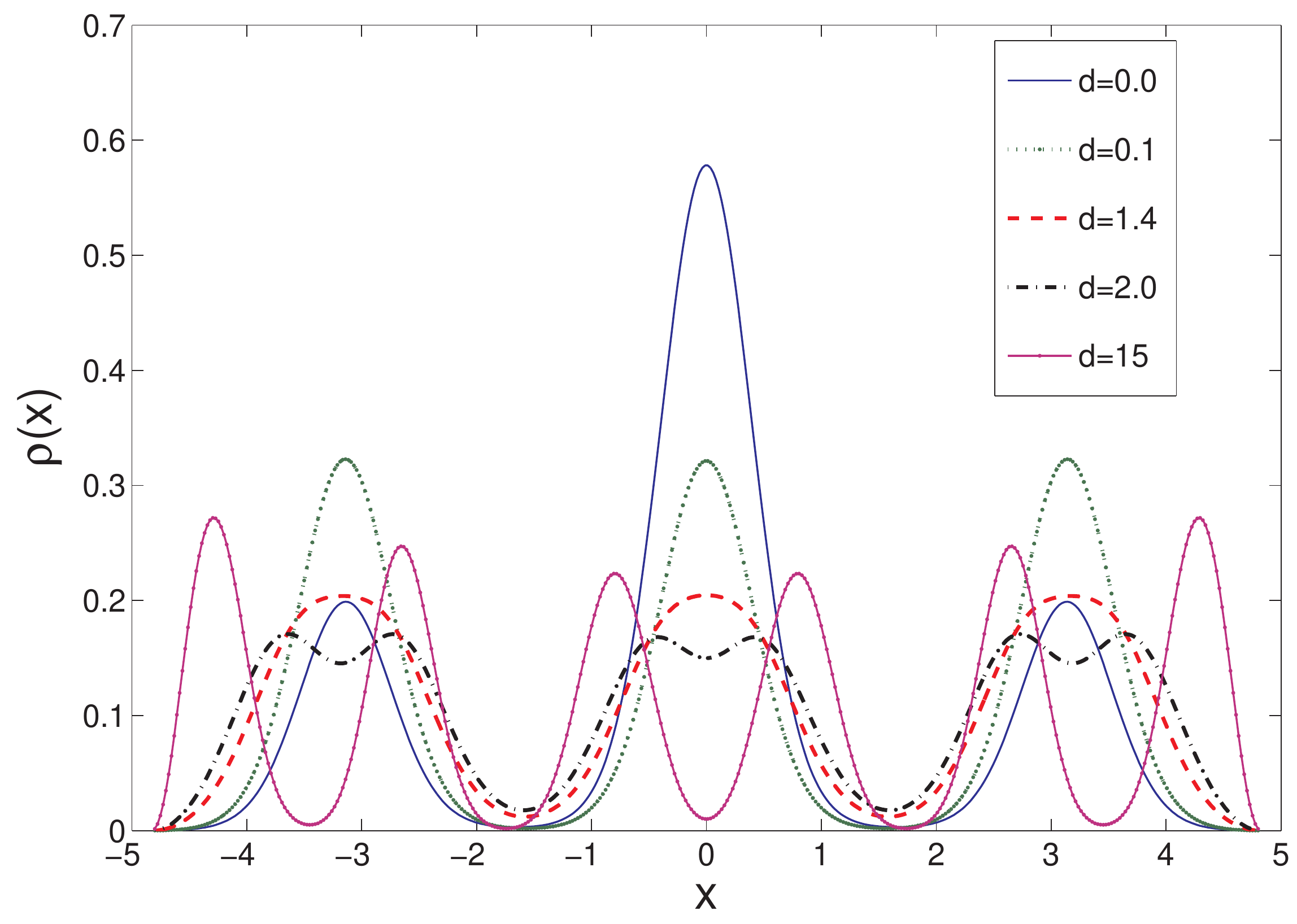}
\end{center}
\caption{(color online) One body density for six bosons in the triple well for different interaction strengths d.\label{cap:den1p_n6}}
\end{figure}

As we observe in Fig. \ref{cap:den1p_n6}  for six atoms (filling factor $\nu=2$) increasing the dipolar interactions leads to an equal population distribution among all the wells. This is due to the tendency to minimize the energy since multiple occupation of a single well is energetically unfavorable for stronger interactions. The system thus forms a Mott-Insulator with each well occupied by two particles. On the top of that, since each well has a double occupancy, the density profile in each site is strongly influenced  by the repulsive dipolar forces and the atoms tend to avoid each other on the same site,  indicating a  localization of the two bosons in each well forming a ``crystal-like state''. This effect is even stronger for dipolar interactions than for contact repulsion \cite{brouzos10} due to their long-range nature. While in the case of contact interacting bosons, the interaction energy saturates at fermionization resulting in the observed  notch at the center, in the case of dipolar bosons there 
is no saturation of the interaction  energy.

\begin{figure}[htb]
\begin{center}
\includegraphics[width=0.32\columnwidth,keepaspectratio]{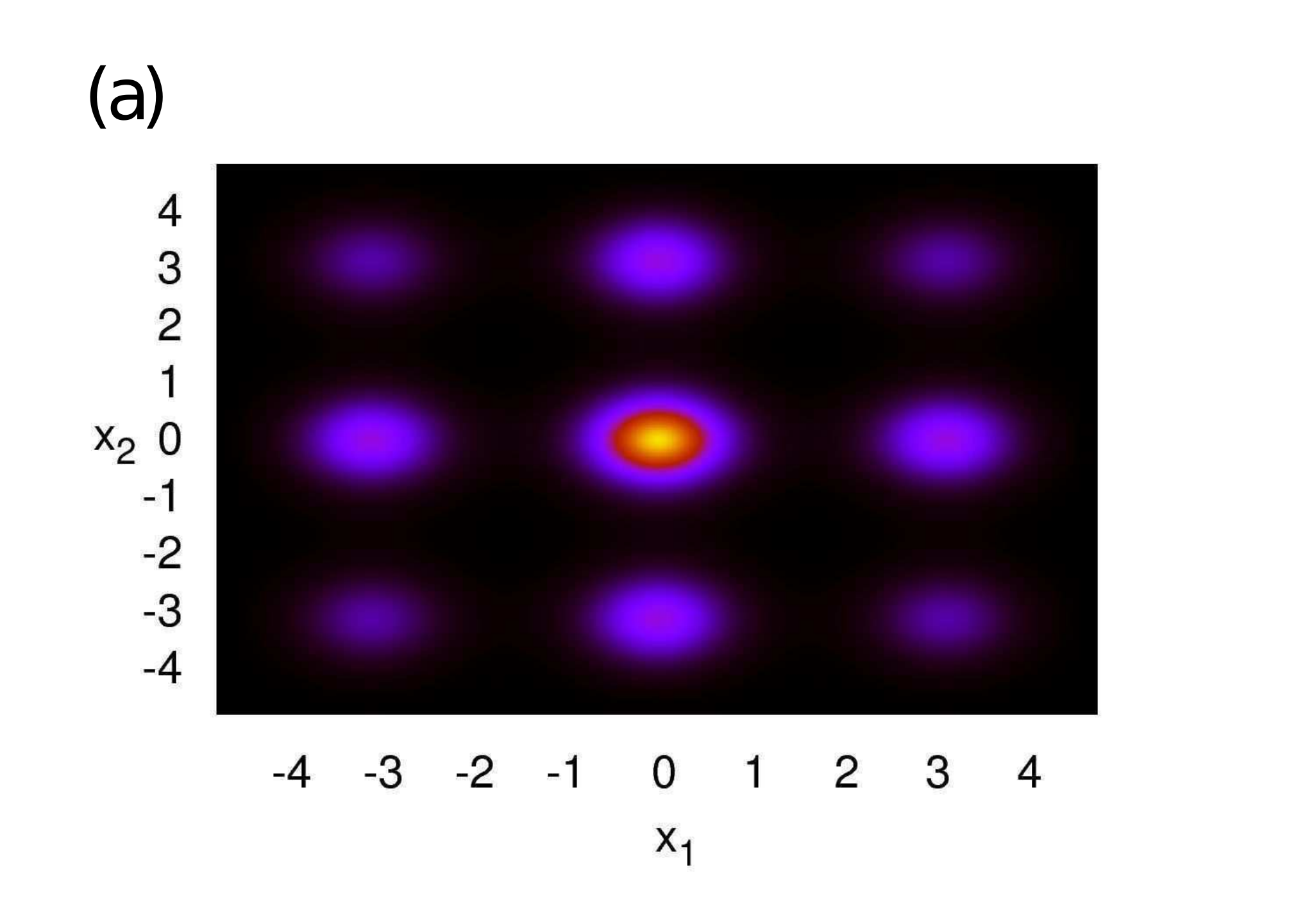}
\includegraphics[width=0.32\columnwidth,keepaspectratio]{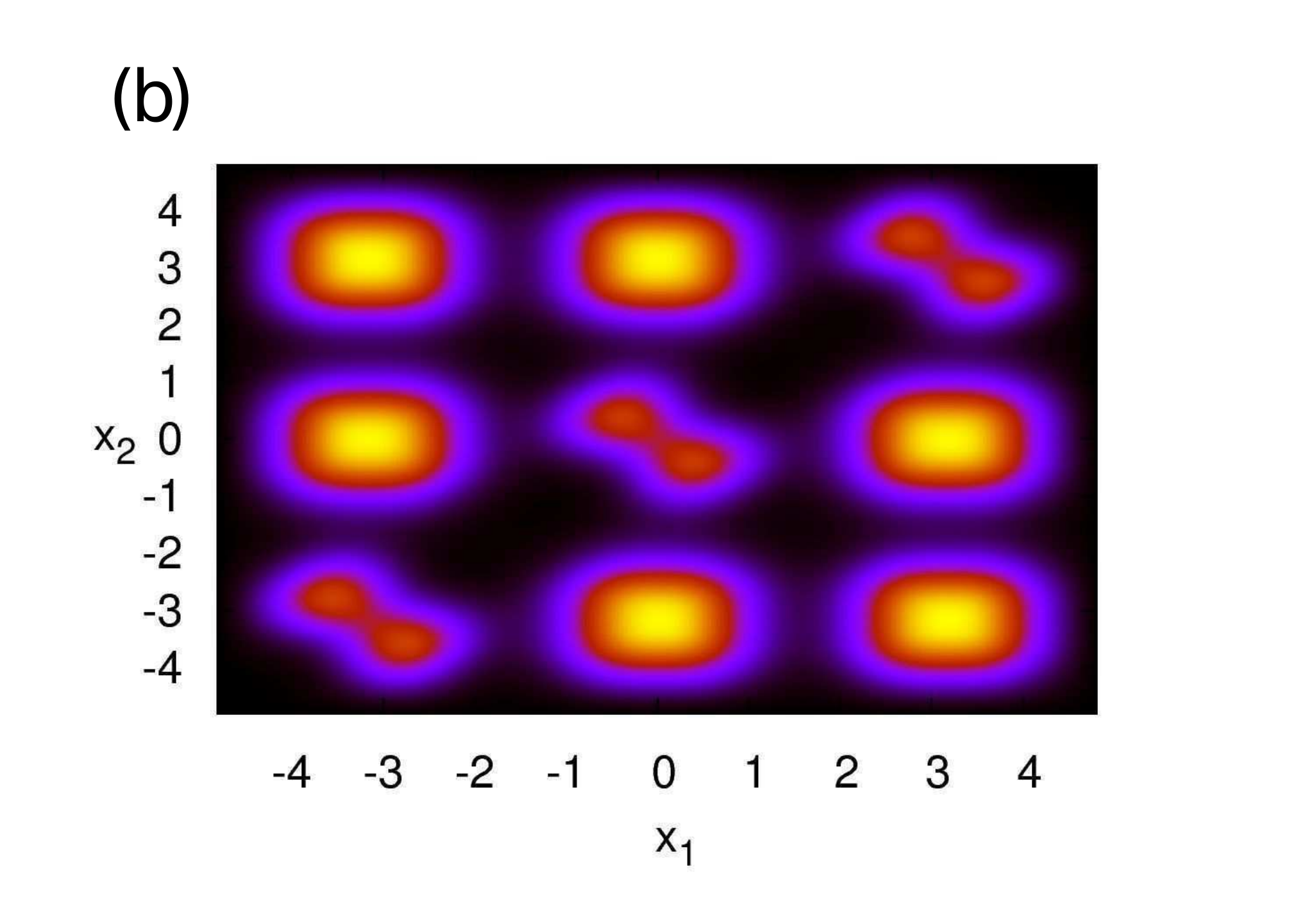}
\includegraphics[width=0.32\columnwidth,keepaspectratio]{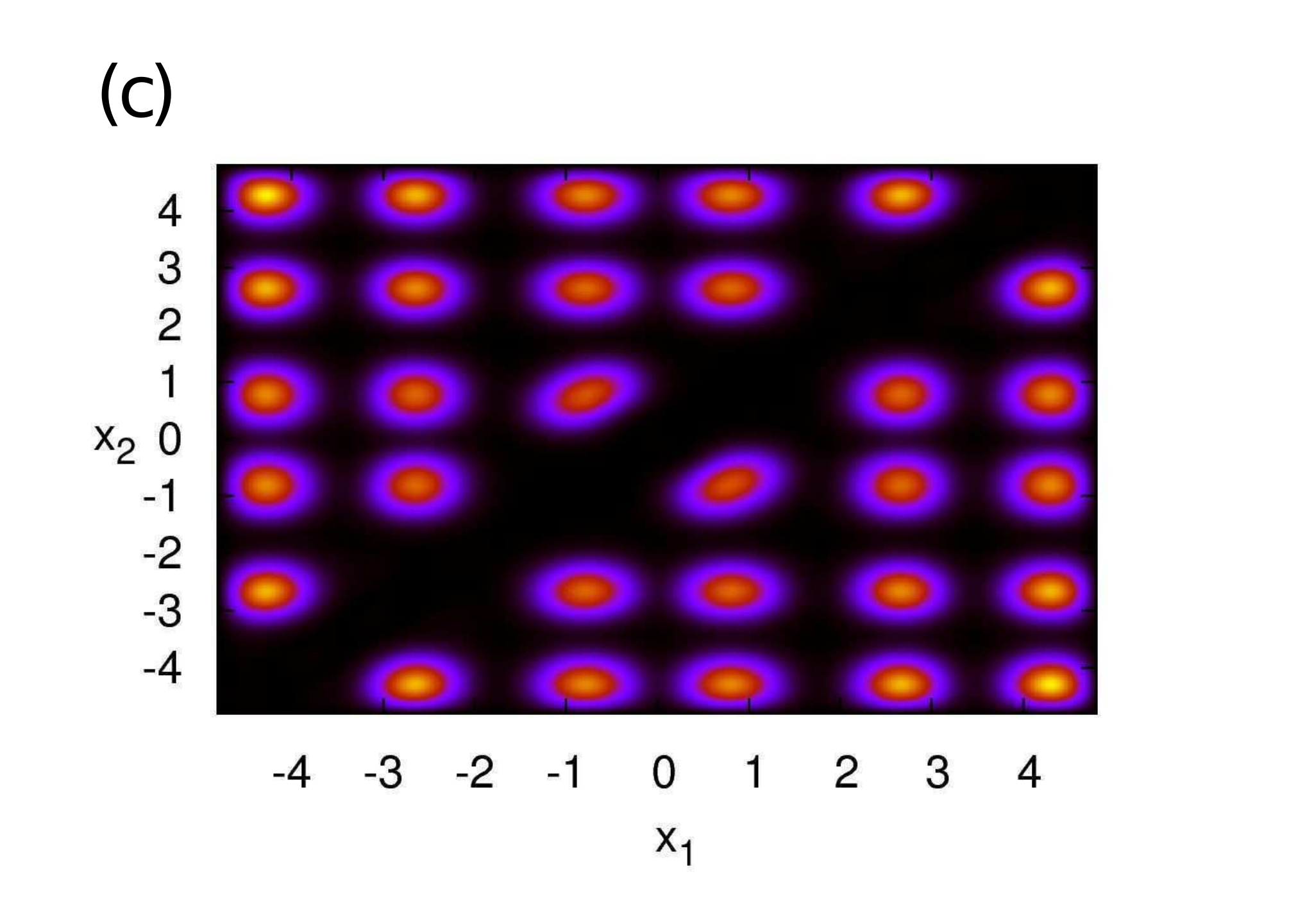}
\end{center}
\caption{(color online): Two body density for six bosons in the triple well for (a) $d=0$ (b) $d=1.4$ (c) $d=15$. \label{cap:den2p_n6}}
\end{figure}

For the corresponding two-body density in Fig. \ref{cap:den2p_n6} we observe a gradual depletion of the diagonal as $d$ increases from $d=0$ and a correlation hole at $x_1=x_2$ can be observed for $d=1.4$ (Fig. \ref{cap:den2p_n6} (b))  This is understandable since the particles in the same well attempt to minimize the energy by avoiding each other and thus the overlap at $x_1=x_2$ is reduced.  For stronger interactions (Fig. \ref{cap:den2p_n6} (c)), the diagonal depletion is accompanied by a fragmentation of the off-diagonal densities revealing an effective 'fermionization' like pattern. 

\subsection{Incommensurate Filling}
\label{sec:incom_fill}

\begin{figure}[htb]
\begin{center}
\includegraphics[width=0.65\columnwidth,keepaspectratio]{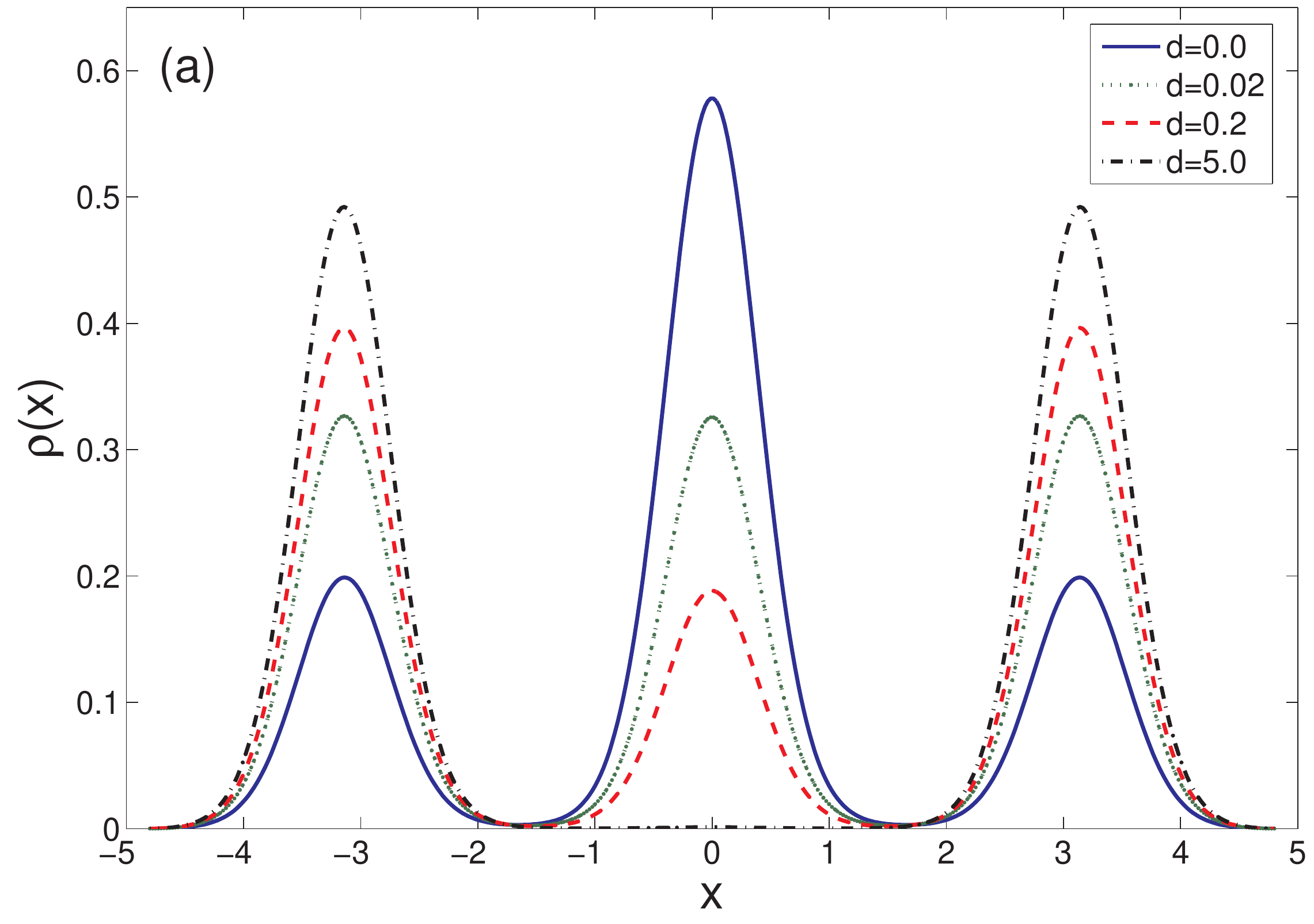}
\includegraphics[width=0.65\columnwidth,keepaspectratio]{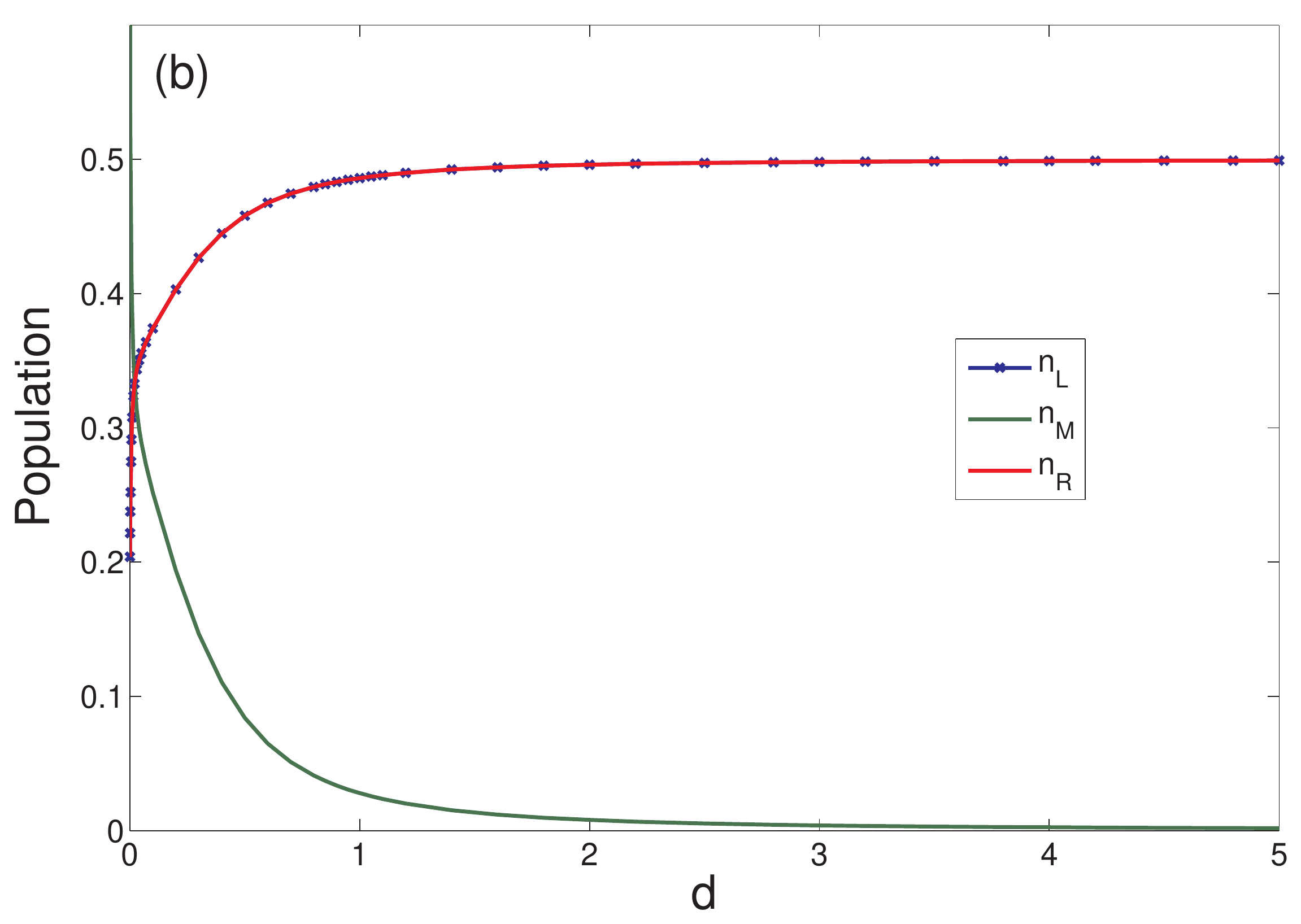}
\end{center}
\caption{(color online) (a) One body density for two bosons in triple well. (b) Variation of population in each well with interaction strength $d$.
\label{cap:den1p_n2}}
\end{figure}

We now study cases with  incommensurate filling, which  show a much stronger effect of the long-range dipolar interactions as well as a stronger dependence on the number of bosons. The most important aspect of incommensurate filling compared to that of the commensurate case is the absence of a true Mott-Insulator state since there is always a superfluid fraction on top of a Mott-Insulator phase. For filling factor $\nu < 1$ (two bosons in a triple well) as the dipolar interaction increases we observe in Fig. \ref{cap:den1p_n2} that the atoms are gradually shifted to the outer wells. The population of the middle well totally vanishes, in contrast to the case of contact interactions \cite{brouzos10}. The long-range dipolar forces act here not only locally or between next neighbours but also between next to next sites. 

\begin{figure}[htb]
\begin{center}
 \includegraphics[width=0.65\columnwidth,keepaspectratio]{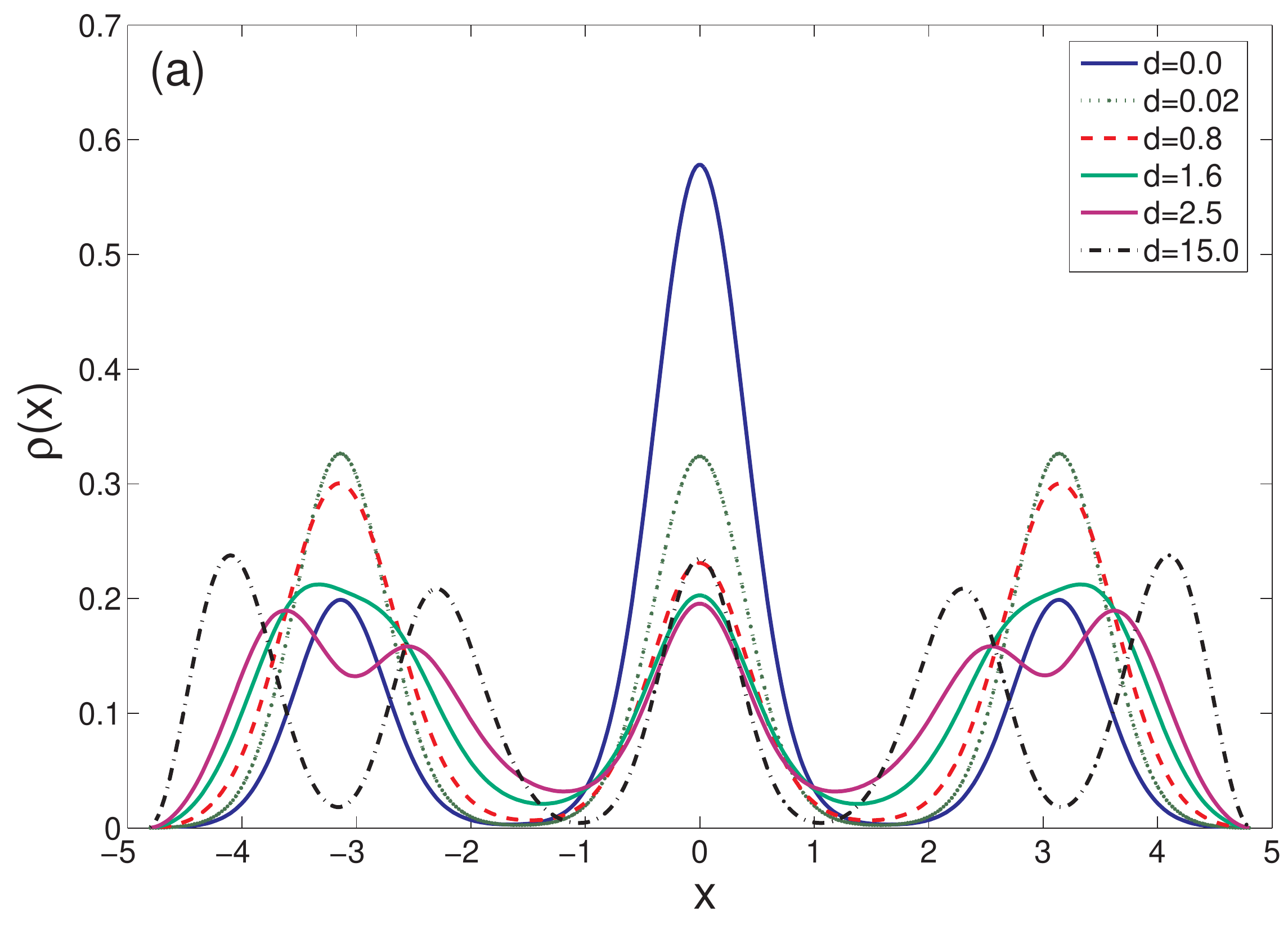}
\includegraphics[width=0.65\columnwidth,keepaspectratio]{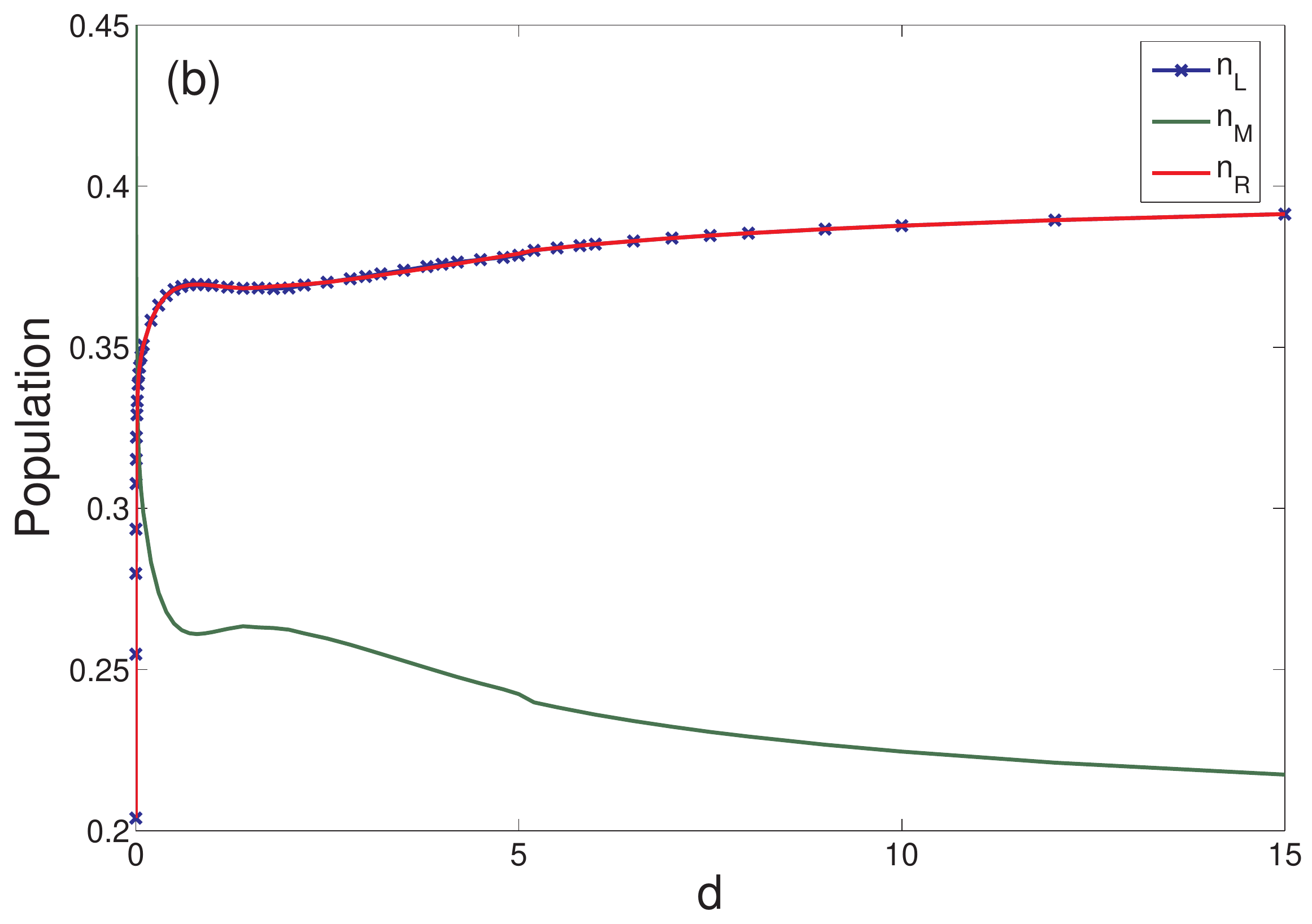}
\end{center}
\caption{(color online) (a) One body density for five bosons in a  triple well for different interaction strengths. (b) Variation of population in each well with varying interaction strength $d$.
\label{cap:den1p_n5}}
\end{figure}

Another strong manifestation of this long-range strong dipolar interaction occurs in the case of a pair of particles on top of a commensurate background (5 bosons in a triple well). For intermediate interactions a plateau region in the population (Fig. 4) occurs followed by a  rearangement of the population with increasing interaction.  We observe a general upward shift of one-body density at the inter-well space accompanied by a broadening of the densities in the outer wells (see $d=1.6$ of Fig.\ref{cap:den1p_n5}(a)). This is a result of the repulsive interaction of the two extra bosons which are pushed to the outer wells. This effect becomes striking for higher interaction strengths where the two extra bosons localize in the outer wells leading to wiggles in the densities in these wells ($d=2.5$ Fig.\ref{cap:den1p_n5}(a)). However, as the interaction increases further, the intra-well repulsive interaction between the bosons in the doubly occupied outer wells  forces them to avoid each other, effectively 
localizing then separately  in the same well. This leads to  the formation of two separate density peaks in the one body density of the outer wells as seen for $d=15$ (Fig.\ref{cap:den1p_n5}(a)). Thus in effect, we obtain a 'crystal-like' state consisting of five separate density peaks with two in the outer wells and a single peak in the middle well. In a sense the strong dipolar force becomes more prominent than the external potential forces, in contrast to the contact interaction which acts only locally.  

\begin{figure}[htb]
\begin{center}
 \includegraphics[width=0.42\columnwidth,keepaspectratio]{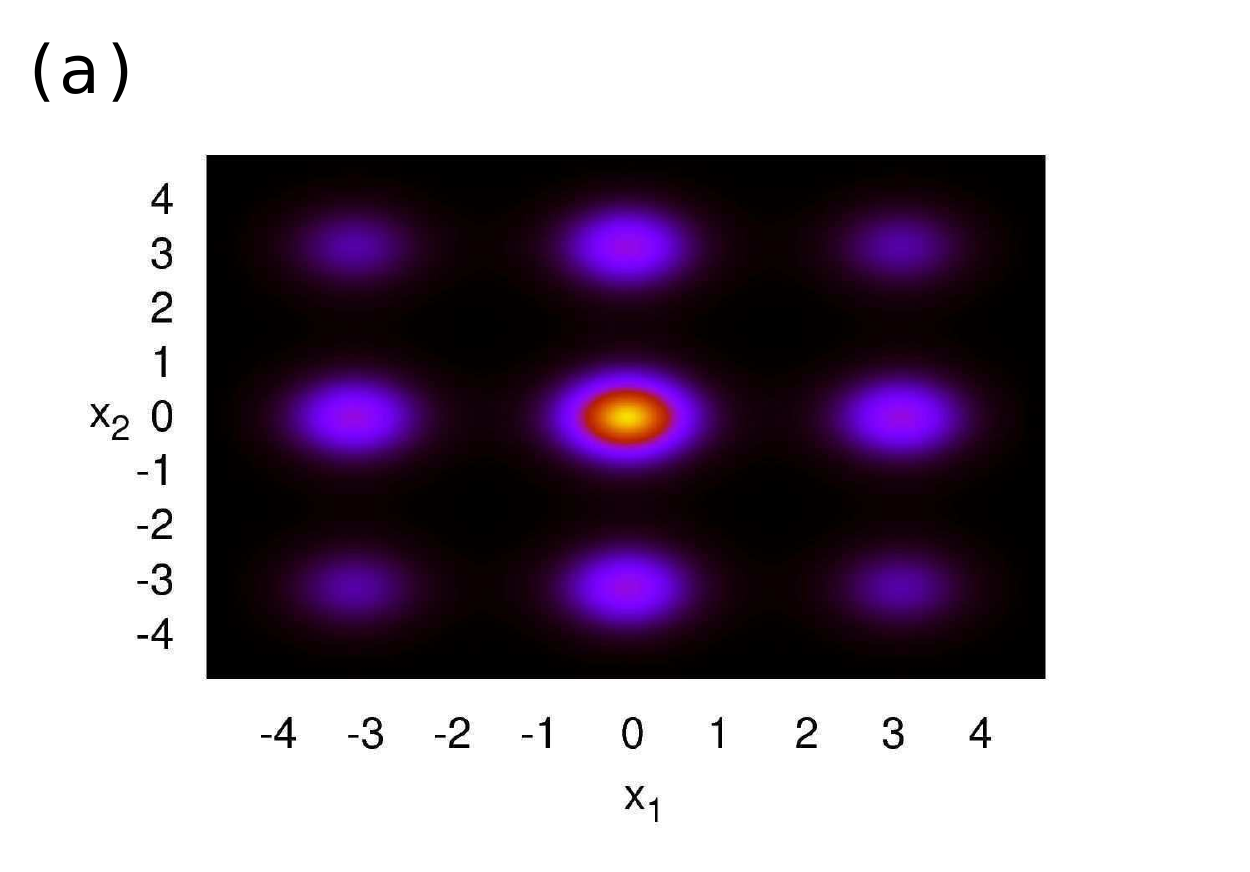}
\includegraphics[width=0.42\columnwidth,keepaspectratio]{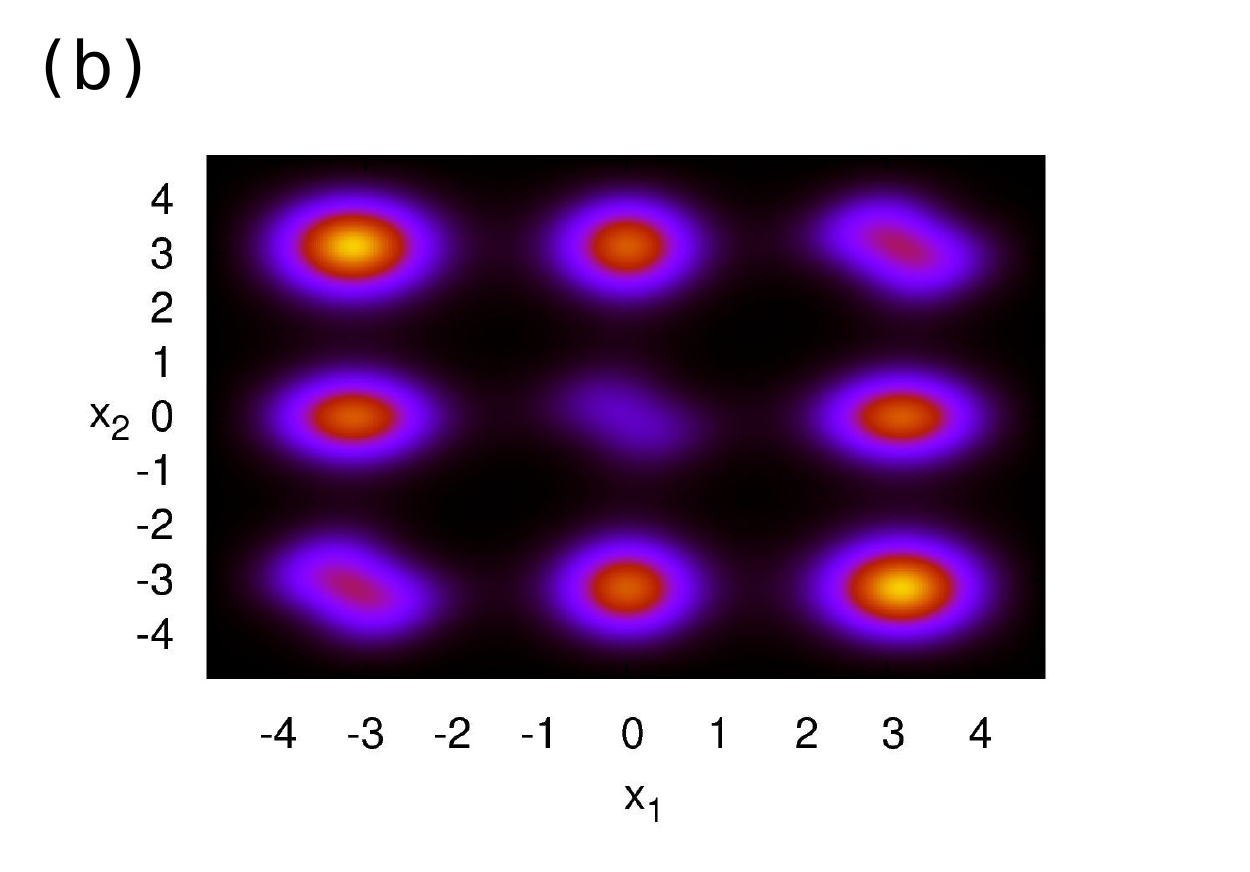}
\includegraphics[width=0.42\columnwidth,keepaspectratio]{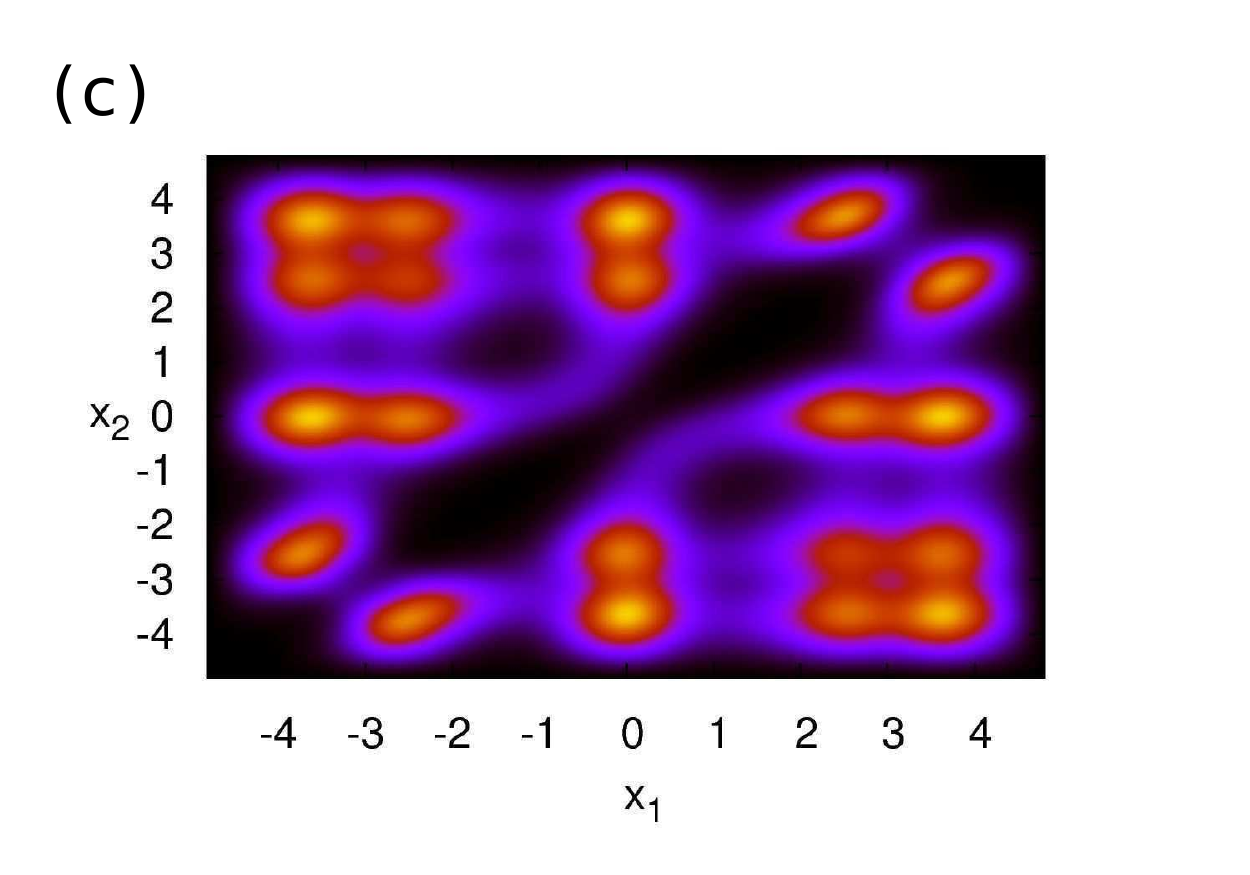}
\includegraphics[width=0.42\columnwidth,keepaspectratio]{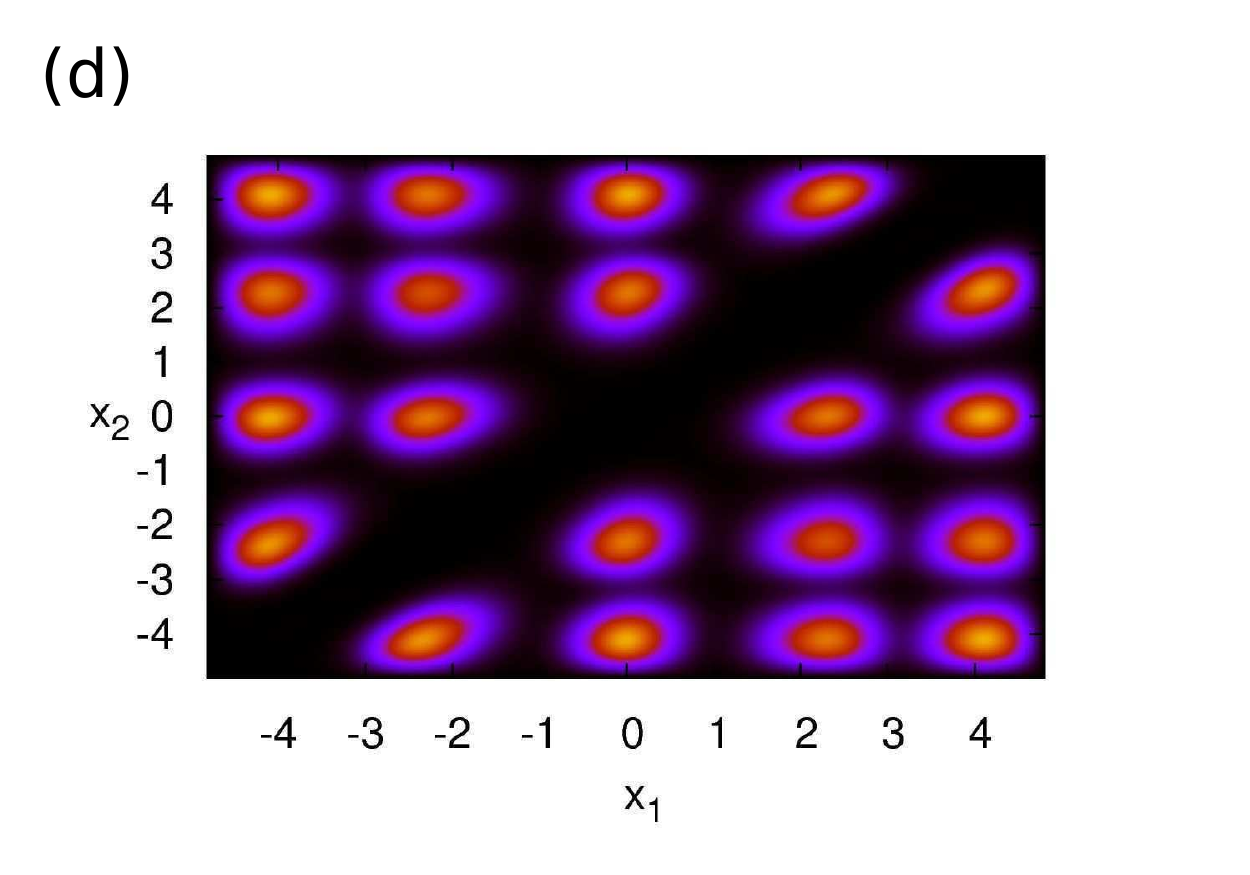}
\end{center}
\caption{(color online) Two  body densities for five bosons in a  triple well for (a) $d=0.0$, (b) $d=0.8$, (c) $d=2.5$, (d) $d=15$.
\label{cap:den2p_n5}}
\end{figure}

Fig. \ref{cap:den2p_n5} presents the two-body densities for different $d$ values. With the transfer of the particles to the outer wells due to the long range repulsion and the depletion of the population in the middle well, the diagonal contribution is strongly reduced for $d=0.8$  (Fig. \ref{cap:den2p_n5}(b)). With further increase in interaction, we observe the formation of the correlation hole which is strongly evident for $d=2.5$ (Fig. \ref{cap:den2p_n5}(c)). Moreover, the off-diagonal density peaks starts to fragment as the two body interactions gain prominence. We see a complete checkerboard formation for $d=15$, with depleted diagonal (Fig.\ref{cap:den2p_n5}(d)) as the strong repulsive interaction creates  strong fragmentation in the off diagonal densities. 

\section{Energy Spectrum}
\label{sec:spec}

As we shall see later dipolar interactions allows for an interesting multi-band tunneling dynamics even for the case of only two bosons. However before we explore the dynamics, let us  first study the energy spectrum for the two particle case. This will guide us and point to the relevant interaction regime and the characteristics of the contributing excited eigenstates.

\begin{figure}[htb]
\begin{center}
 \includegraphics[width=0.75\columnwidth,keepaspectratio]{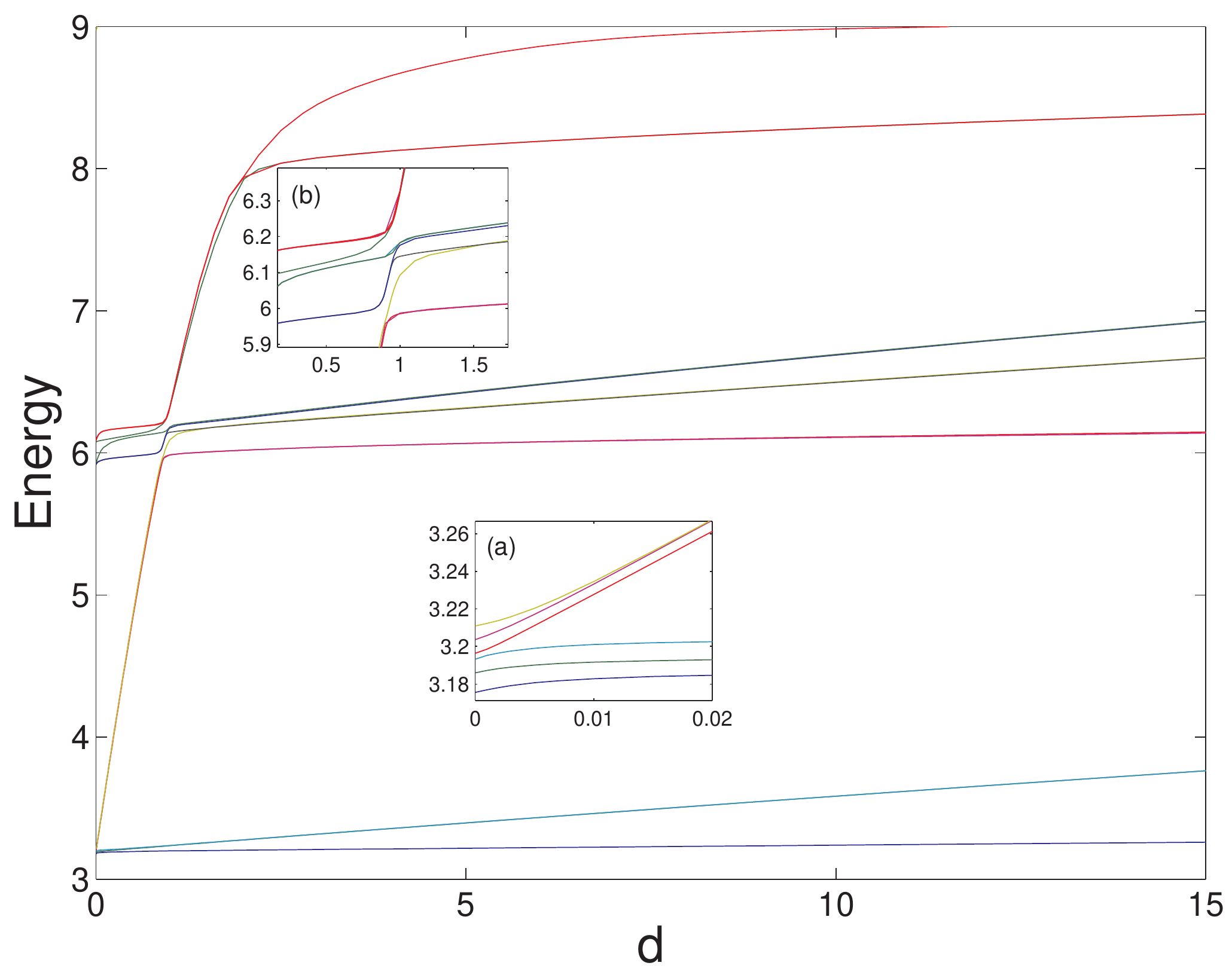}
\end{center}
\caption{(color online) Two boson energy spectrum as a function of $d$. Inset: (a) Magnified lowest band energy spectrum for small interaction strength. (b) Magnified energy spectrum at the vicinity of the avoided crossings. 
\label{cap:energy_spec}}
\end{figure}

\subsection{Lowest band states}

In Fig. \ref{cap:energy_spec}, we present the energy spectrum for two bosons as a function of the interaction strength $d$. For weak interactions, the states in the lowest band ( first six states) are separated into two groups of triplets. The lowest three are composed of the number-states $|1,0,1\rangle$, $|1,1,0\rangle$ and $|0,1,1\rangle$ (the standard number state notation means the population of each well eg. the first vector denotes the state where one boson is in the left and one in the right well). However, because of the long-range nature of the dipolar interaction, these states are not equivalent energetically ( the $|1,0,1\rangle$ state possessing a  lower energy) and thus the lowest triplet breaks with increasing dipolar interaction.

The upper triplet is composed of the doubly occupied number-states $|2,0,0\rangle$, $|0,0,2\rangle$ and $|0,2,0\rangle$ with the state $|0,2,0\rangle$ possessing a slightly lower energy because of the hard-wall boundary conditions.
These states have a much stronger dependence with $d$ as a result  of the double occupancy in the wells. As the interaction increases, the highest two states join together forming a doublet separating from the lower energy state (Fig. \ref{cap:energy_spec} inset (a)). Unlike the case of contact interactions, here the energy does not saturate in the fermionization regime but continues to increase as the interaction strength increases. This has as an important consequence for the interplay with the higher band states.

\subsection{Higher band states}

The increase in energy of the lowest band states with increasing $d$ leads to avoided crossings  with the energies of the higher band states and to corresponding resonances. The  higher band states involving primarily singly occupied number-states  $|1^1,1^0,0\rangle$, $|1^0,1^1,0\rangle$, $|0,1^1,1^0\rangle$, $|0,1^0,1^1\rangle$ and  $|1^1,0,1^0\rangle$ and $|1^0,0,1^1\rangle$ (where the superscript now refers to the band index) have a much weaker dependence of the energy on $d$ compared to the doubly occupied number states of the lower band and hence  we observe with increasing $d$  avoided crossings in the energy spectrum among these states (see Fig. \ref{cap:energy_spec} inset (b)). We note that since the states $|1^n,0,1^m\rangle$ are energetically detuned from the states $|1^n,1^m,0\rangle$ and $|0,1^n,1^m\rangle$, there are actually two resonances with the lowest band states. These crossings lead to an interesting inter-band tunneling dynamics for two particles as we will see next in Sec.\ref{sec:dyn}.

In general the two-body spectrum is here much richer than in the case of contact interaction, since a saturation at fermionization is not forseen for dipolar forces, and thus the long range interaction mixes lower and higher states via crossings and anticrossings.

\section{Quantum Dynamics near resonances}
\label{sec:dyn}

The avoided crossings occurring in the two-body energy  spectrum open the possibility for an interesting multi-band quantum dynamics. We thus  focus within our study of the tunneling dynamics of two bosons  on the interaction regime in the vicinity of the avoided crossings ($d=0.8-1.2$).

\begin{figure}[htb]

\begin{center}
 
\includegraphics[width=0.40\columnwidth,keepaspectratio]{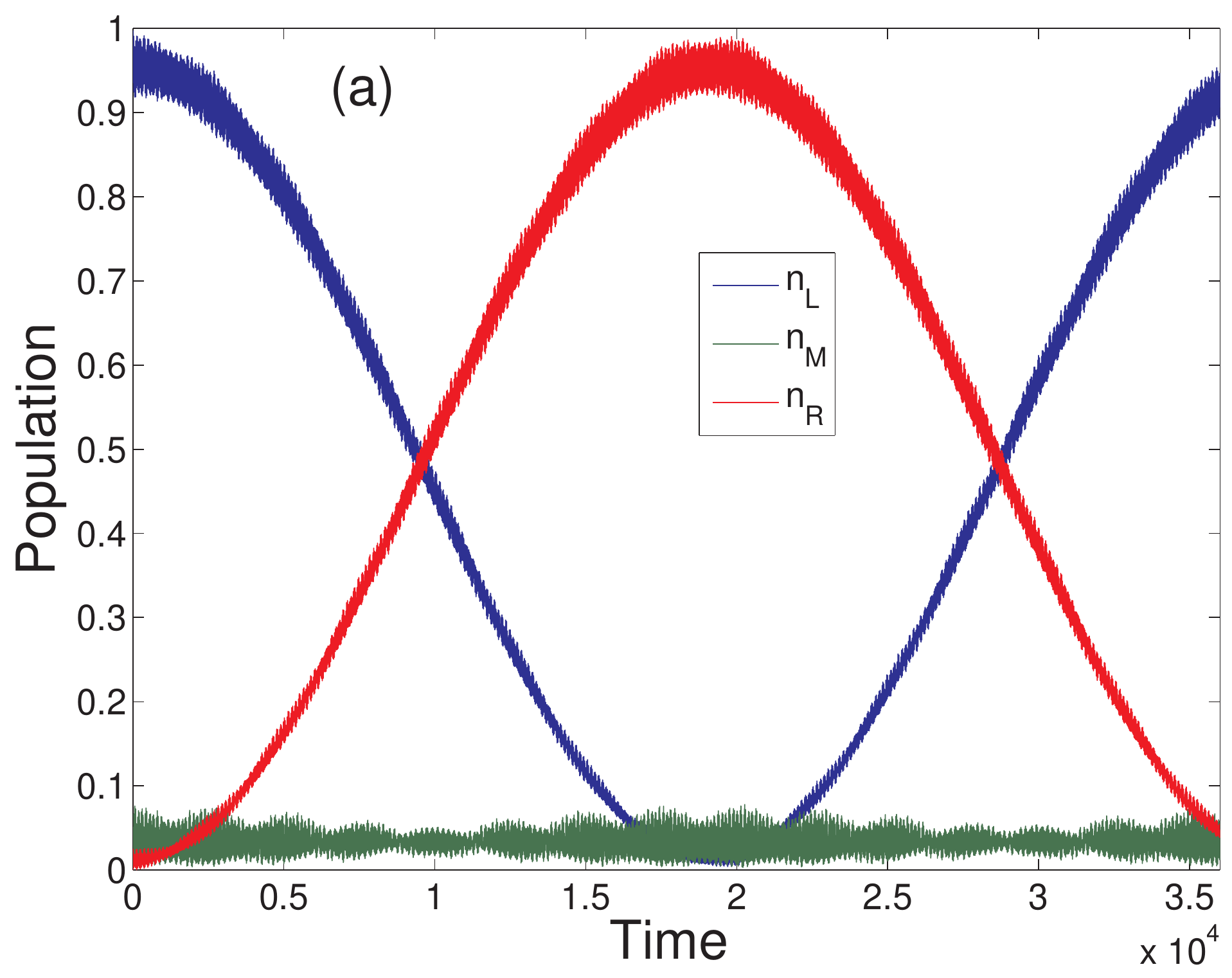}
\includegraphics[width=0.42\columnwidth,keepaspectratio]{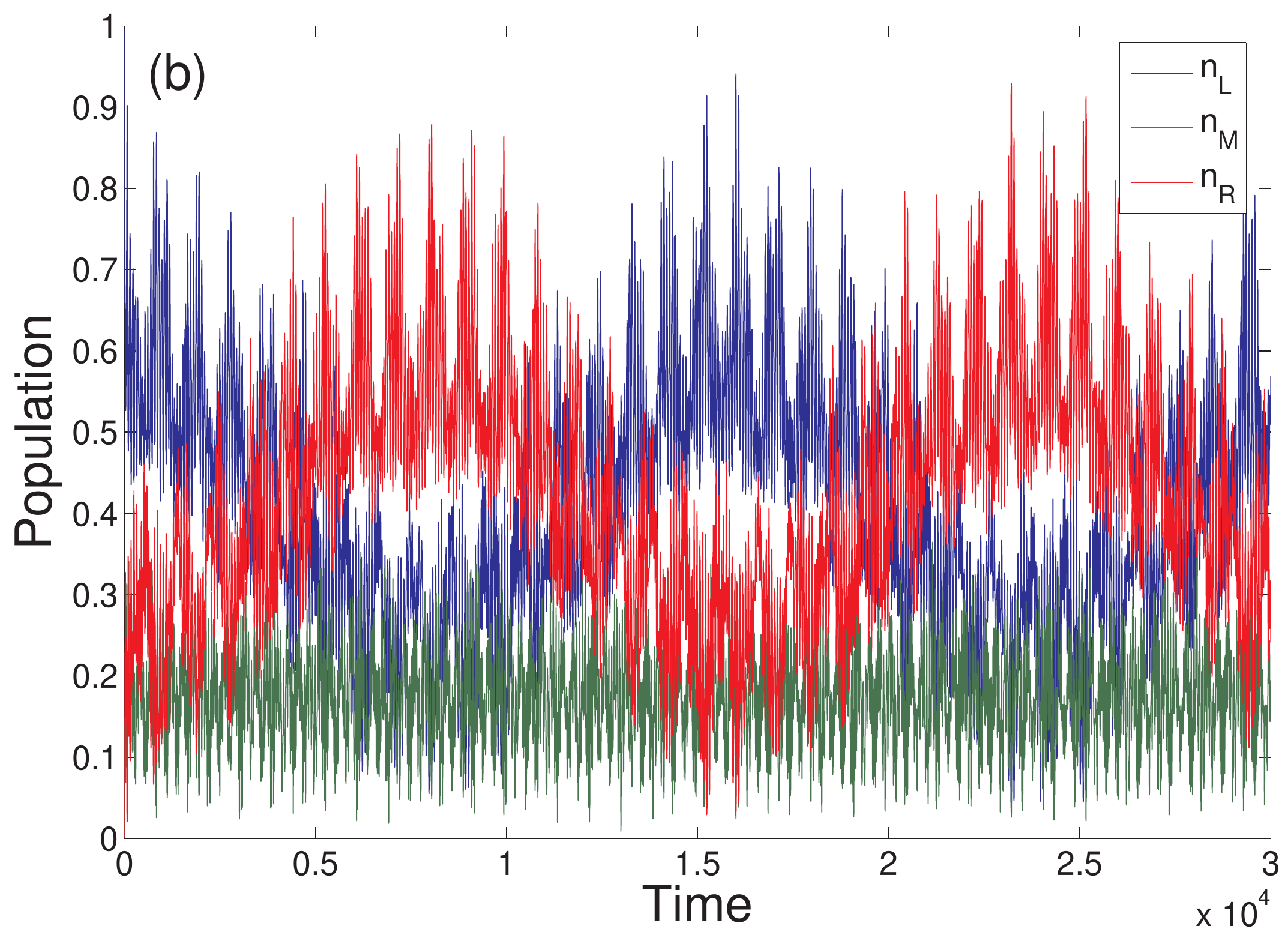}
\includegraphics[width=0.42\columnwidth,keepaspectratio]{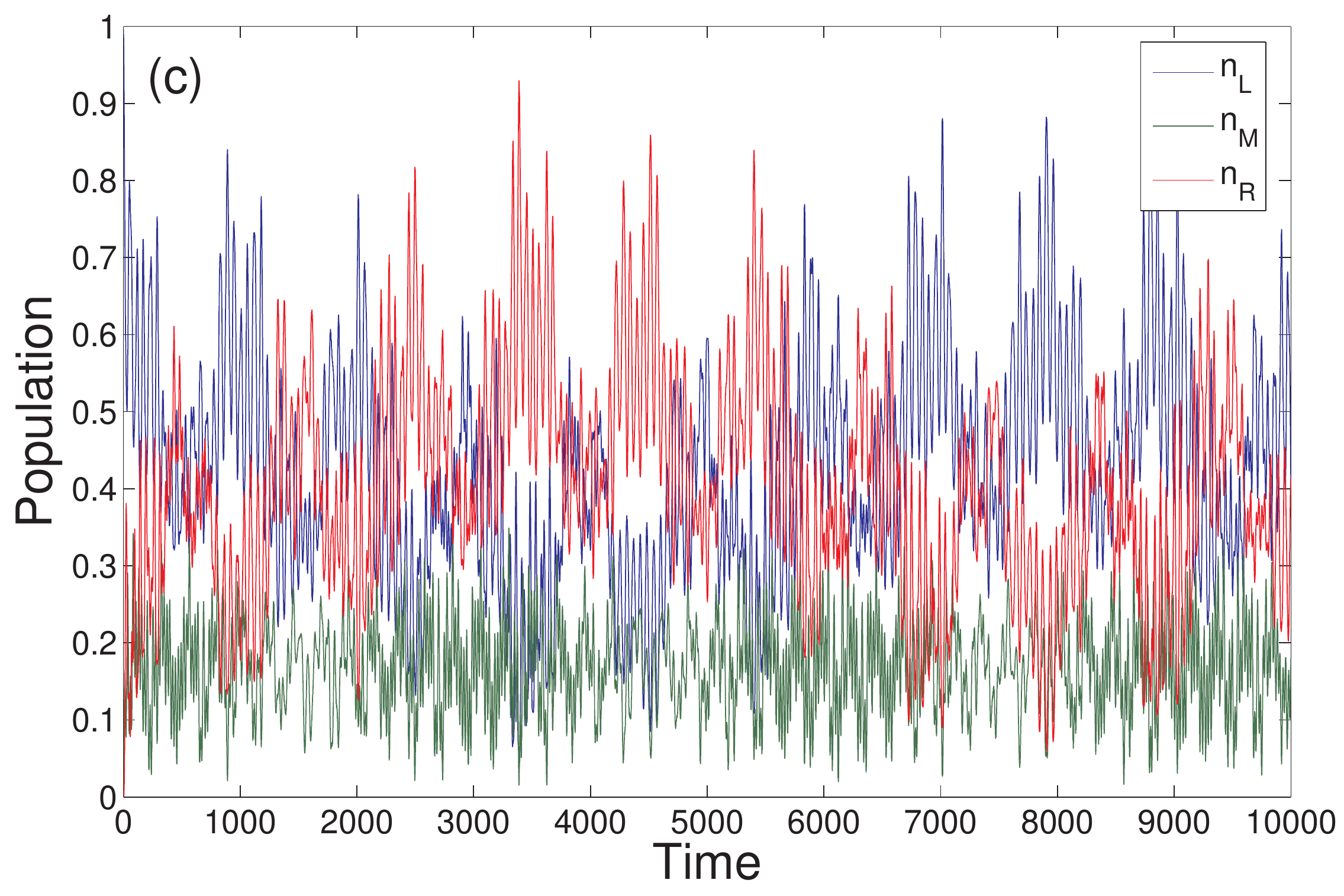}
\includegraphics[width=0.42\columnwidth,keepaspectratio]{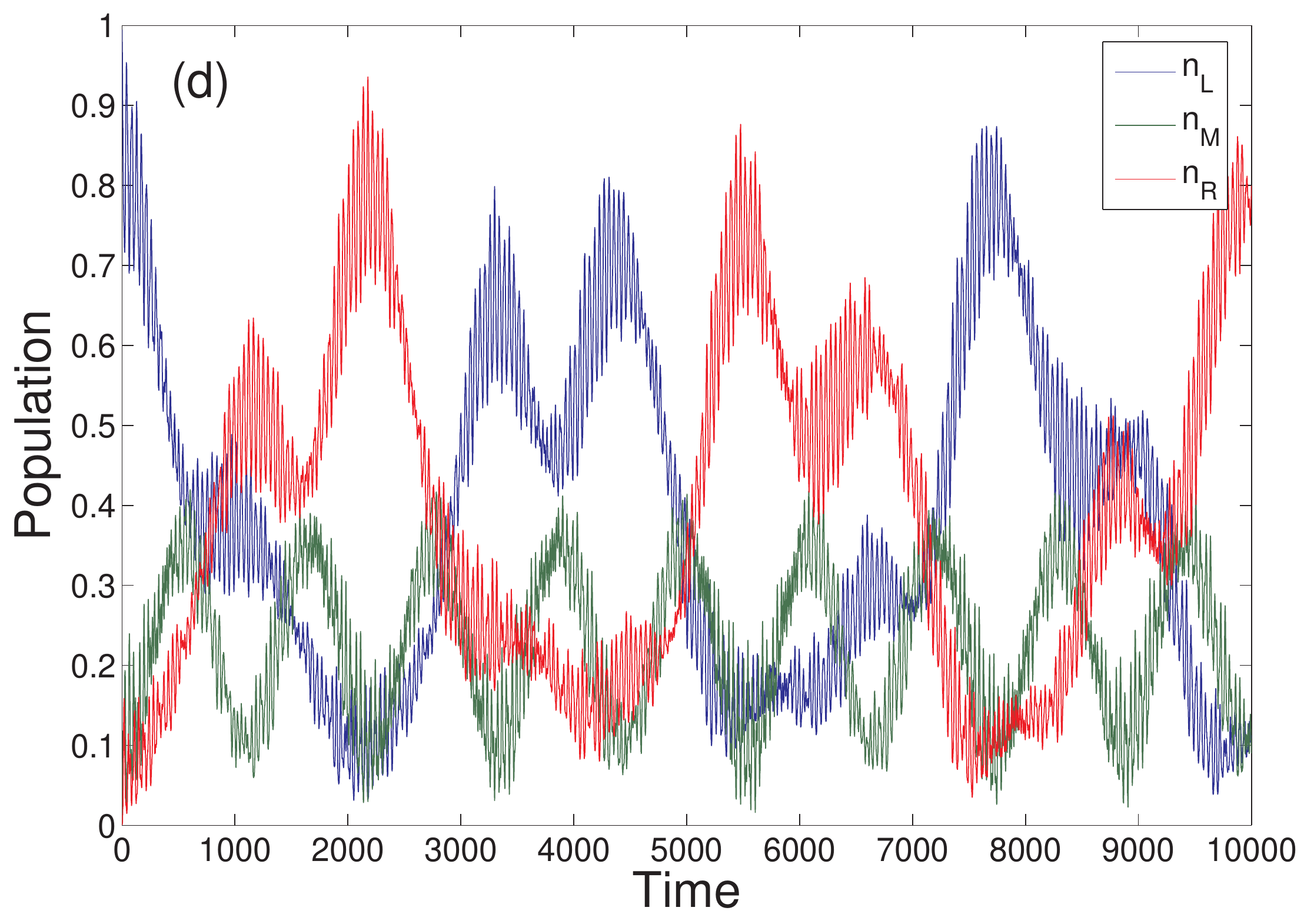}

\end{center}

\caption{(color online) Time evolution of the population density in the left, middle and right well for (a) $d=0.8$, (b) $d=0.9$, (c) $d=0.95$, (d) $d=1.0$
\label{cap:dyn}}
\end{figure}

\begin{figure}[htb]

\begin{center}
 
\includegraphics[width=0.42\columnwidth,keepaspectratio]{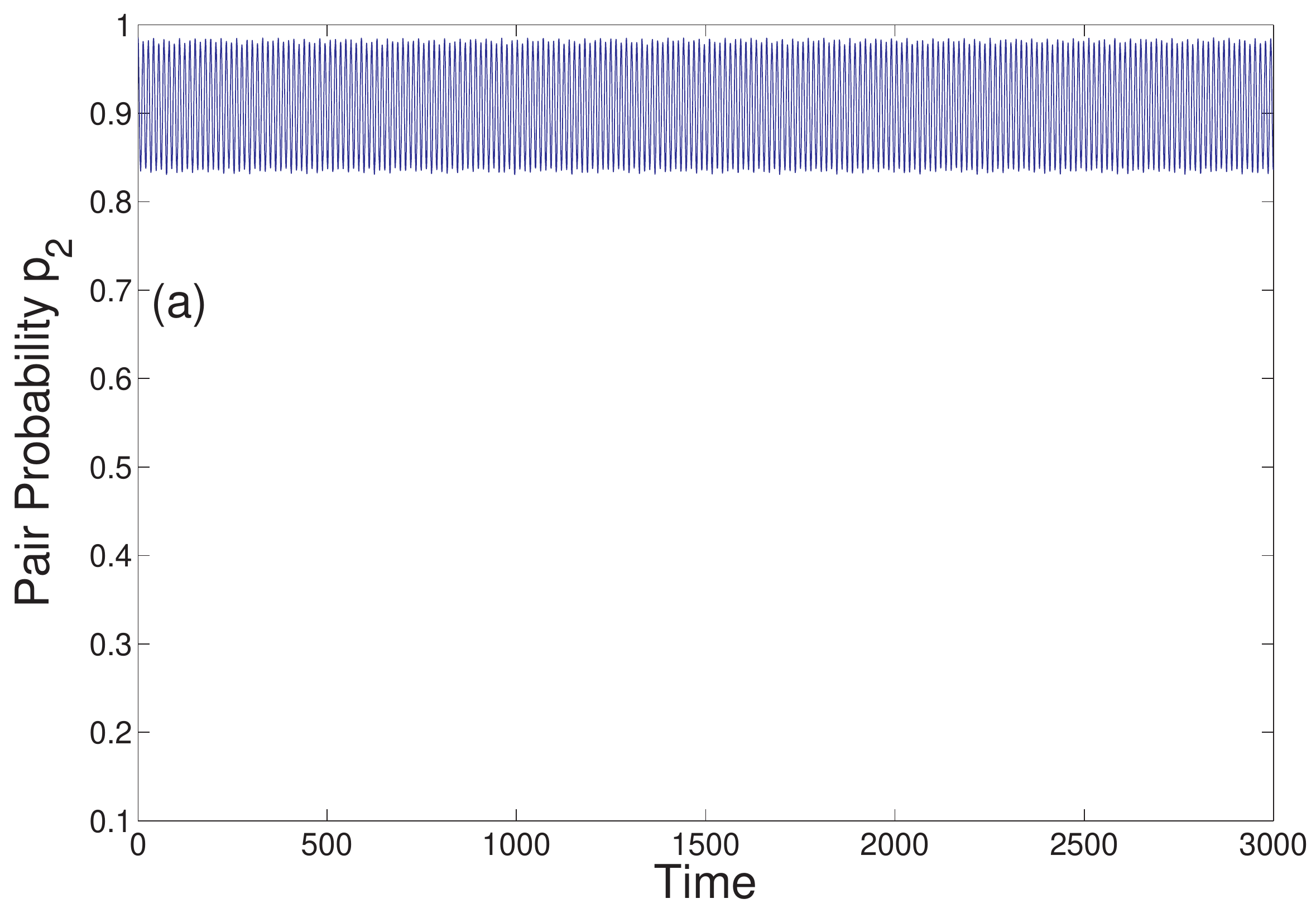}
\includegraphics[width=0.42\columnwidth,keepaspectratio]{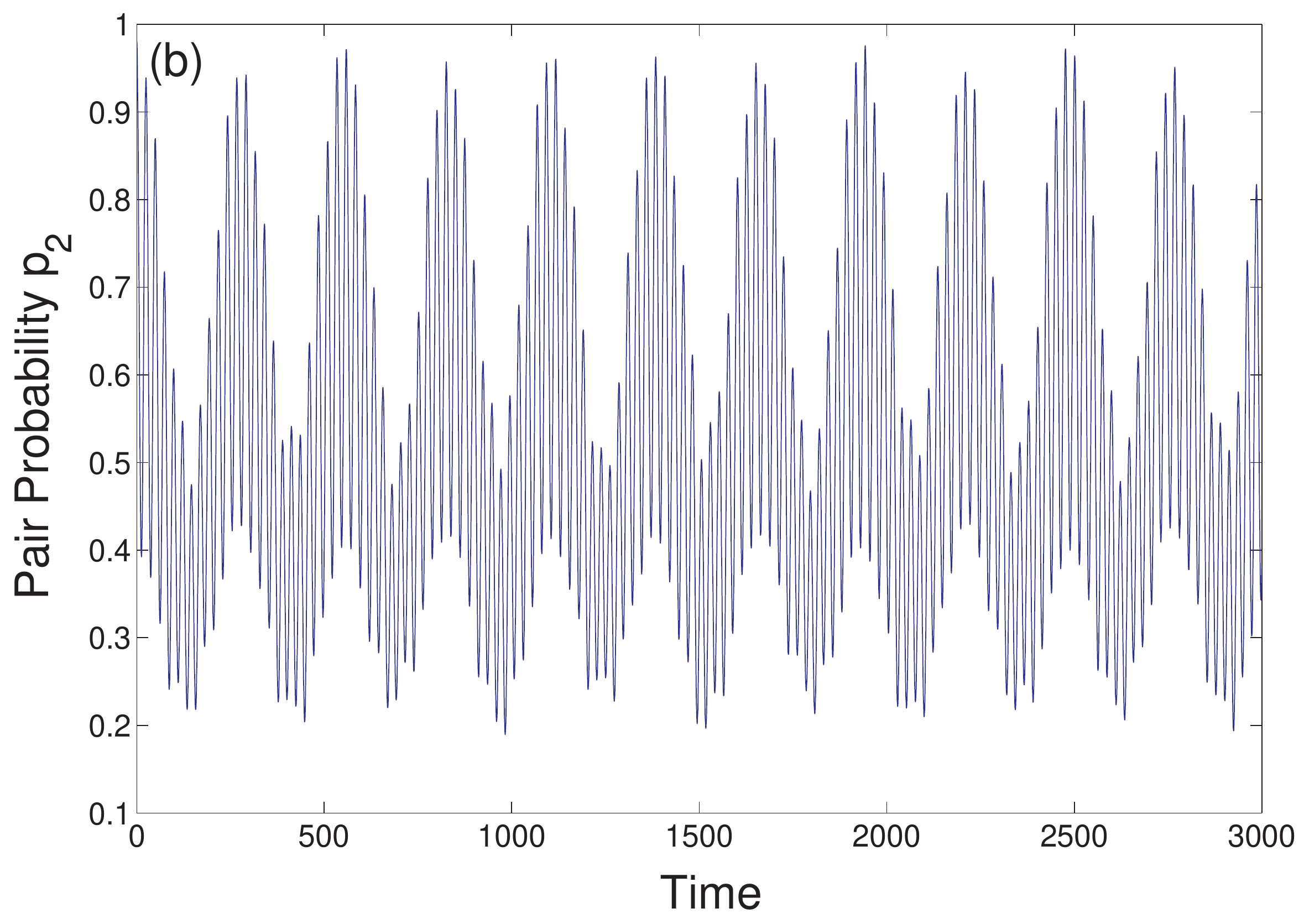}
\includegraphics[width=0.42\columnwidth,keepaspectratio]{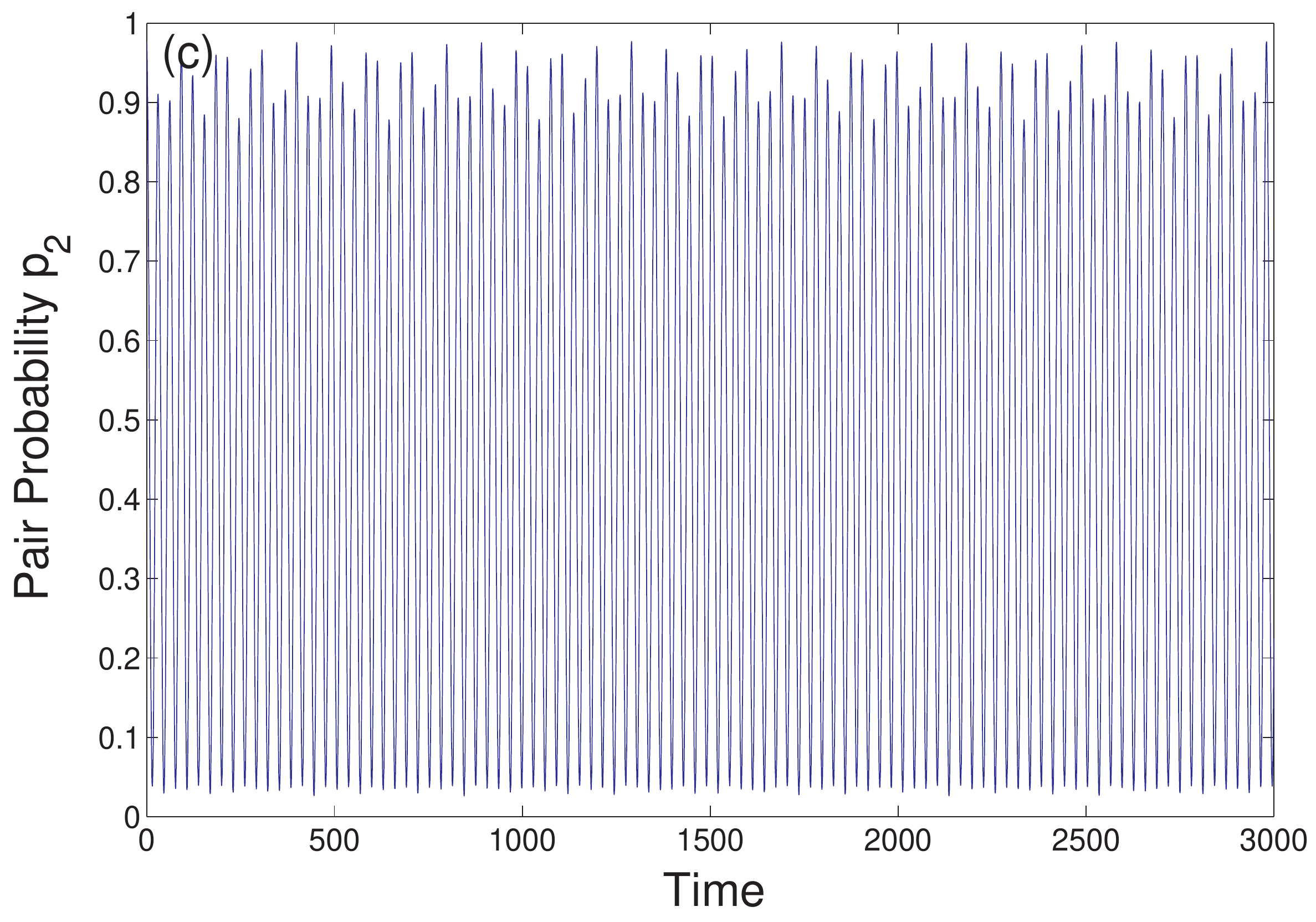}
\includegraphics[width=0.42\columnwidth,keepaspectratio]{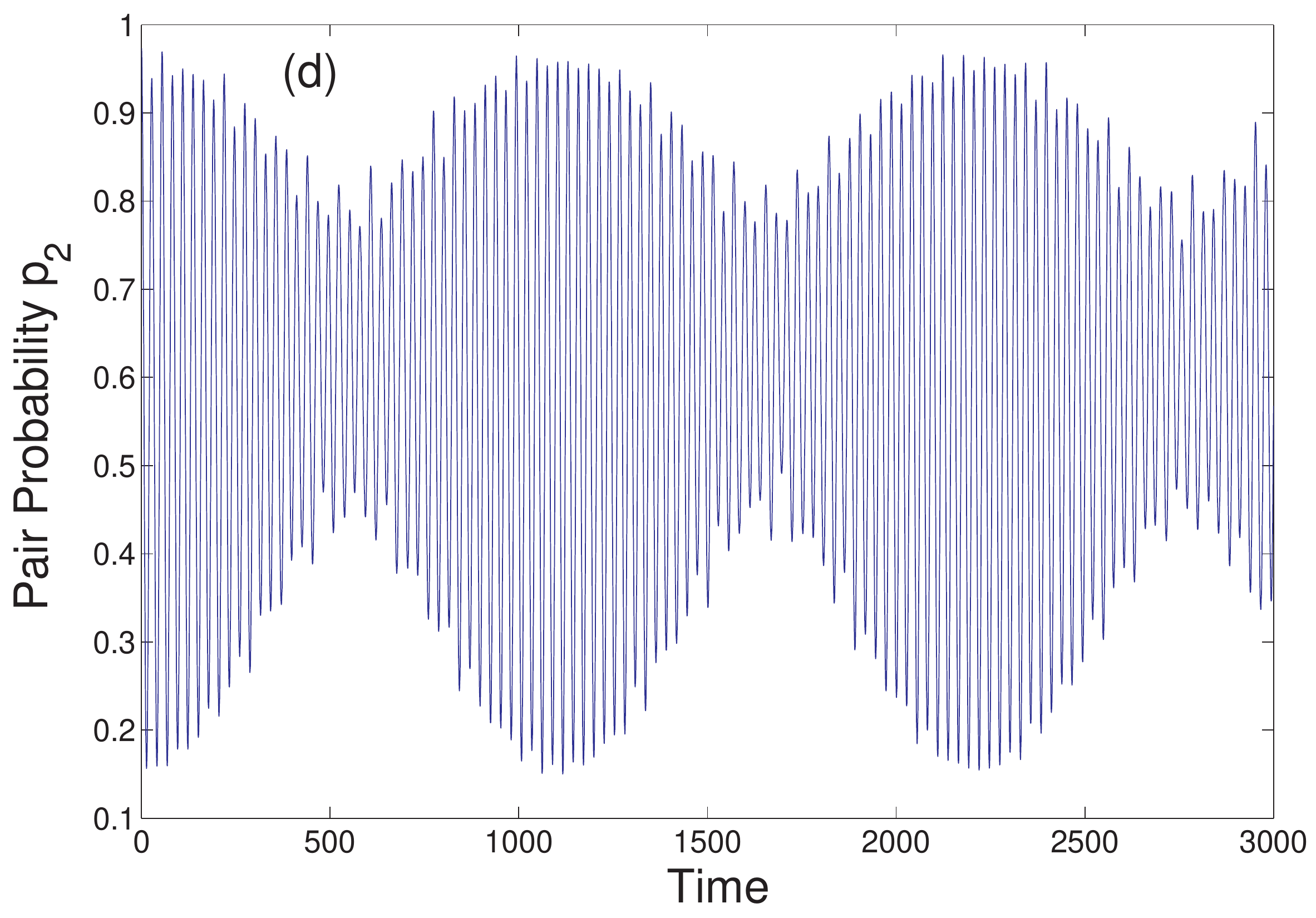}

\end{center}

\caption{(color online) Time evolution of the pair probability for the same values of $d$ as in Fig. 7.
\label{cap:npair}}
\end{figure}

We first consider an initial setup for which both bosons are localized in the left well. 
In Fig. \ref{cap:dyn} the evolution of the population of  the three different wells as a function of time is shown for different values of $d$. For $d=0.8$, which is just below the resonance, we find a uniform dynamical evolution of the populations possessing a long time period (Fig. \ref{cap:dyn}(a)). The initial population in the left well  tunnels to the right well with a very long period, while the middle well remains almost unpopulated. This is a demonstration of pair-tunneling arising  primarily from energy resonance of two doubly occupied number-states namely $|2,0,0\rangle$ and $|0,0,2\rangle$. This can be confirmed by studying the pair-probability, that is, the probability of finding two atoms in the same well. Fig. \ref{cap:npair}(a) shows the evolution of the pair-probability and we observe that it remains and oscillates close to unity - implying that the bosons tunnel as pairs and there is minimal single particle tunneling. This repulsively bound pair phenomenon (or doublon) is also seen and 
observed experimentally \cite{pair_tunneling} for contact interactions, we verify here that it is also a possible mechanism for dipolar forces.  

For $d=0.9$ the resonant region (in the spectrum) is encountered, and we observe a tunneling dynamics containing a combination of many frequencies (Fig. \ref{cap:dyn}(b)), i.e.,  many eigenstates contribute equally to the dynamics. For long time scales, we observe a dominant tunneling from the left to the right well as indicated by their respective populations. However now, there is a substantial population in the middle well which oscillates around $20 \%$. This long time oscillation is modulated by combinations of fast high frequency oscillations. The mechanism of the tunneling could be understood as follows: The tunneling primarily involves transfer from $|2,0,0\rangle$ to $|0,0,2\rangle$ via the intermediate number-states  $|1^n,1^m,0\rangle$, $|1^n,0,1^m\rangle$ and $|0,1^n,1^m\rangle$ (where $m$ and $n$ indicate the band index). Away from the avoided crossings, the intermediate states are energetically far detuned from the initial and final states resulting in very long tunneling timescale and minimal 
contributions of the intermediate number-state. Near the avoided crossings, the intermediate singly occupied number state of the excited band are energetically in resonance with the initial and final states. This result in both faster tunneling and significant contributions of the intermediate states resulting in both enhanced single particle tunneling and middle well population. The predominance of single particle tunneling is reflected in the evolution of the pair-probability (Fig. \ref{cap:npair}(b)) which oscillates between unity and $0.2$ implying that in the course of the tunneling it is highly probable to find two particles in separate wells, thereby breaking the doublon dynamics. 

Increasing the interaction ($d=0.95$) leads to further shortening of the tunneling period with even larger modulations due to the participating different frequencies making the pattern of the tunneling highly irregular as seen in Fig. \ref{cap:dyn}(c). Here, we observe the absence of a complete left to right tunneling even for long timescales. Instead, especially for shorter timescales, the single particle tunneling dominates. About 2/3 of the population tunnels from the left well  to the right one with a portion remaining in the middle well, while the rest remains in the left well. The mechanism for this tunneling is primarily the resonance between the states $|2,0,0\rangle$ and $|1^n,0,1^m\rangle \pm |0,1^n,1^m\rangle$. This is reflected in the pair-probability (Fig. \ref{cap:npair}(c)) which shows oscillations between unity and near zero, thus displaying the oscillation between a pair-state and a singly occupied state.

A different mechanism is encountered as the interaction is further increased to $d=1.0$.   In  Fig. \ref{cap:dyn}(d) we observe a much larger population in the middle well in the course of the dynamics.  A complete tunneling from the left to the right well occurs for long timescales. Moreover, the numerous large frequency modulations seen in the previous case are here reduced although the periodic nature of long-time tunneling is not prominently displayed. 
To understand this we note that unlike that of contact interaction, for dipolar bosons, the state $|1^n,0,1^m\rangle$ becomes energetically offset from the other singly occupied number state as a consequence of the long-range nature of the interaction. Hence, a second resonance is encountered where the contribution of the state $|1^n,0,1^m\rangle$ becomes minimal. The mechanism is primarily a resonance between the $|2,0,0\rangle$ and $|0,1^n,1^m\rangle$ states leading to a population transfer to both the middle and right well. Also involved in this resonance, we  have the tunneling channel between the two doubly occupied number-states $|2,0,0\rangle$ and $|0,0,2\rangle$ which is valid especially for longer timescales. The pair-probability (Fig. \ref{cap:npair}(d)) captures this  mechanism, showing a two mode oscillation. A very fast oscillation, ranging from unity to 0.2 implying the breaking of pair and a longer oscillation envelope constitute the dynamics.
Increasing $d$ further, we go beyond the crossing region thus resulting in a reoccurence of pair tunneling (not shown here).

\section{Conclusion}
\label{sec:conclusion}

We have investigated the ground state properties and quantum dynamics of repulsively interacting dipolar boson in a triple well trap. The crossover of the ground state from the weak to strong interaction regime has been covered for different filling factors focusing primarily on dipolar effects which occur due to the long-range nature of this interaction potential and differ from that of the pure contact interaction case. In the case of commensurate filling, apart from the usual Superfluid-Mott insulator transition, for very  strong interaction, we observed intra-site localization and extremely strong fragmentation. For incommensurate filling, additionally we find a general tendency for avoiding the middle well since the long range nature of the dipolar force pushes the atoms to the outer wells while the exact ground state configuration depends strongly on the exact number of particles considered. For strong dipolar forces a crystal like phase appears which is largely unaffected by the external potential. 

The two-boson energy spectrum shows avoided crossings between the energy levels of different bands. In the region of these crossings, we observe an inter-band tunneling dynamics arising from the energetic resonances between the localized number-states corresponding to different bands. This regime exhibits different tunneling mechanisms ranging from complete pair tunneling to single particle higher band dynamics. This study of the static and dynamic properties of ultracold dipolar boson for small bosonic ensembles can serve as a very good starting-point for our understanding of the strong interaction regime in larger systems. It might also help to develop schemes for creation of novel entangled few-body states and their controlled quantum dynamics and transport.  

\acknowledgments{ B. C. acknowledges support from the Landesexzellenzinitiative Hamburg, which is financed by the Science and Research Foundation, Hamburg and supported by the Joachim Herz Stiftung. L. C. and P. S. gratefully acknowledge financial support by the Deutsche Forschungsgemeinschaft (DFG) in the framework of the collaborative research center SFB 925.}

\end{document}